\documentclass{IEEEtaes}

\usepackage{xcolor,array,amsthm}
\usepackage{graphicx}
\usepackage{soul}
\usepackage{orcidlink}
\usepackage{amsmath}
\usepackage{tikz} 
\usepackage{tabularx}    
\usepackage{multicol}    
\usepackage{multirow}    
\usepackage{longtable}   
\usepackage{booktabs}    
\usepackage{ltablex}
\usepackage{setspace}
\usepackage[export]{adjustbox}
\usepackage{subcaption}  
\usepackage[capitalize,noabbrev]{cleveref}
\usepackage{soul}
\usepackage{url}
\usepackage{algorithm}
\usepackage{algpseudocode}
\usepackage{amssymb}

\pubyear{2026}

\begin{document}

\title{Enhanced input stacking for non-square MIMO modal identification of aeronautical structures via Fast and Relaxed Vector Fitting} 

\author{BEATRICE E. BAURET MARTÍNEZ}
\affil{Universidad Carlos III de Madrid, Leganés, Madrid, Spain} 

\author{GABRIELE DESSENA \orcidlink{0000-0001-7394-9303}} 
\affil{Universidad Carlos III de Madrid, Leganés, Madrid, Spain}

\author{MARCO CIVERA \orcidlink{0000-0003-0414-7440}} 
\affil{Politecnico di Torino, Turin, Piedmont, Italy}

\author{OSCAR E. BONILLA-MANRIQUE \orcidlink{0000-0003-0541-8310}}
\affil{Universidad Carlos III de Madrid, Leganés, Madrid, Spain}


\receiveddate{
This work has been supported by the Madrid Government (\emph{Comunidad de Madrid} – Spain) under the Multiannual Agreement with the Universidad Carlos III de Madrid (\href{https://researchportal.uc3m.es/display/act564873}{IA\_aCTRl-CM-UC3M}). The first author's stay at Politecnico di Torino during her graduate thesis was supported by the Erasmus+ Traineeship Programme of the European Union. The first author is supported by grant JDC2024-055593-I funded by MICIU/AEI/10.13039/501100011033 and by ESF+.
{The third author is supported by the Centro Nazionale per la Mobilità Sostenibile (MOST – Sustainable Mobility Centre), Spoke 7 (Cooperative Connected and Automated Mobility and Smart Infrastructures), Work Package 4 (Resilience of Networks, Structural Health Monitoring, and Asset Management).}}
\corresp{{\itshape (Primary corresponding author: M. Civera; Corresponding author: G. Dessena)}. The second and third authors contributed equally to this work.}


\authoraddress{B.E. Bauret Martínez and G. Dessena are with the Department of Aerospace Engineering at the Universidad Carlos III de Madrid, Av.da de la Universidad 30, Leganés, 28911, Madrid, Spain. (e-mails: \href{mailto:100510852@alumnos.uc3m.es}{100510852@alumnos.uc3m.es} and \href{mailto:gdessena@ing.uc3m.es}{gdessena@ing.uc3m.es}). M. Civera is with the Department of Structural, Geotechnical and Building Engineering at the Politecnico di Torino, Corso Duca degli Abruzzi 24, 10129, Turin, Italy (e-mail: \href{mailto:marco.civera@polito.it}{marco.civera@polito.it}). O.E. Bonilla-Manrique is with the Electronic Technology Department at the Universidad Carlos III de Madrid, Av.da de la Universidad 30, Leganés, 28911, Madrid, Spain. (e-mail: \href{mailto:obonilla@ing.uc3m.es}{obonilla@ing.uc3m.es}).}

\editor{The MIMO Fast and Relaxed Vector Fitting implementation (Data file 1 -- A tutorial for Fast and Relaxed Vector Fitting for modal analysis) used in this work is openly available from the Zenodo repository at [DOI to be reserved] and the identification results and numerical dataset (Data file 2 -- Data supporting: Enhanced input stacking for non-square MIMO modal identification of aeronautical structures via Fast and Relaxed Vector Fitting) supporting this study are openly available from the Zenodo repository at [DOI to be reserved].
In addition, this study used third-party experimental data (Data file 3 -- BAE T1A Hawk Full Structure Modal Test) made available at [\url{https://orda.shef.ac.uk/articles/dataset/BAE_T1A_Hawk_Full_Structure_Modal_Test/24948549}].
Data files 1 and 2 are available under the terms of the [GNU General Public License v3.0 (GPL 3.0)].}
\supplementary{\textit{Authorship credit statement}: Conceptualisation, BE.BM., G.D., and M.C.; methodology, BE.BM., G.D., and M.C.; software, BE.BM. and M.C.; validation, BE.BM., G.D., and M.C.; formal analysis, BE.BM.; investigation, BE.BM.; resources, G.D., M.C., and OE.BM.; data curation, BE.BM.; writing---original draft preparation, BE.BM.; writing---review and editing, BE.BM, G.D., M.C., and OE.BM.; visualisation, BE.BM, G.D., and M.C.; supervision, G.D. and M.C.; project administration, G.D. and M.C.; funding acquisition, G.D., M.C., and OE.BM.. All authors have read and agreed to the published version of the manuscript.}

\markboth{BAURET MARTÍNEZ ET AL.}{Enhanced input stacking for non-square MIMO identification of aerostructures via FRVF}
\maketitle

\begin{abstract}
Fast and Relaxed Vector Fitting (FRVF) is a frequency-domain system identification approach that has been widely adopted in electrical system modelling, while its application to mechanical systems has remained relatively unexplored. In this work, FRVF is reformulated for the identification of structural modal parameters of an aircraft based on Ground Vibration Test (GVT) data within a Multi-Input Multi-Output (MIMO) framework. The proposed procedure consists of three stages: (i) rational approximation of frequency response functions via an enhanced input-stacking strategy, (ii) identification of system poles from the resulting rational model, and (iii) estimation of modal parameters from the extracted poles and associated residues. The methodology is first numerically validated on a MIMO beam model, with particular emphasis on accuracy and robustness under increasing measurement noise. Subsequently, experimental validation is conducted using GVT data from the BAE Systems Hawk T1A aircraft. The results obtained demonstrate a level of performance comparable to that achieved by existing methods. Overall, the extended MIMO formulation of FRVF exhibits high accuracy and strong robustness to measurement noise, highlighting its suitability for application in GVT-based modal analysis.

\end{abstract}

\begin{IEEEkeywords}
Aeronautical Structures, Ground Vibration Test, System Identification, Modal Analysis, Multi-Input Multi-Output, Enhanced Input Stacking, Vector Fitting.
\end{IEEEkeywords}

\section{INTRODUCTION}
A precise representation of aircraft structural dynamics is a fundamental requirement for the initial certification of airworthiness \cite{DeFlorio2011}. Within aeronautics, Ground Vibration Testing (GVT) constitutes the experimental basis for the identification of natural frequencies (either $f_n$, in Hz, or $\omega_n=2\pi_n$, in rad/s), damping ratios ($\zeta_n$), and mode shapes ($\mathbf{\phi}_n$), which are essential for finite element model updating \cite{Dessena2024a} and for the prediction of aeroelastic instabilities, such as flutter \cite{Dessena2025_damping_flutter}.

The extraction of these modal parameters from measured data is typically addressed through System Identification (SI) methodologies. Although Single-Input Multiple-Output (SIMO) formulations are often adequate, the increasing structural complexity of modern aerospace vehicles requires the adoption of Multi-Input Multi-Output (MIMO) frameworks to accurately represent coupled and closely spaced vibration modes. Classical identification techniques, including the Least Squares Complex Exponential (LSCE) method \cite{Spitznogle1971}, frequently exhibit limited robustness and degraded performance in the presence of measurement noise, requiring careful windowing to mitigate spectral leakage.  LSCE estimates the modal parameters by fitting sums of complex exponentials to free-decay responses derived from impulse response functions. In contrast, frequency-domain techniques, such as the improved Loewner Framework (iLF) \cite{Dessena2024}, operate directly in the frequency-domain, thus naturally avoiding explicit reconstruction of impulse responses. The iLF constructs reduced-order models from Frequency Response Functions (FRFs) using Loewner matrices and tangential interpolation, offering reliable broadband MIMO identification. However, the computational cost of the classical implementation of the Loewner Framework can become significant for large-scale datasets \cite{Palitta2022}.

By comparison, Fast and Relaxed Vector Fitting (FRVF), originally introduced as Vector Fitting (VF) in the context of electrical network modelling \cite{Gustavsen1999a} and later extended with its fast \cite{Deschrijver2008} and relaxed versions \cite{Vector2006}, provides a computationally efficient and robust frequency-domain alternative for system identification. In the present study, FRVF is reformulated for its application to general MIMO structural systems, thus accommodating non-square input–output configurations commonly encountered in full-scale aircraft GVT campaigns. This overcomes the well-known restrictions of the current FRVF implementation to square systems only \cite{GrivetTalocia2015}, i.e., applications in which the number of input and output channels coincide. Indeed, while this constraint is rarely restrictive in electrical circuit testing (where FRVF originates from), it becomes critical in experimental modal analysis for the identification of mechanical systems, where the number of measured response channels typically far exceeds the available excitation locations.

Through the use of an enhanced input-stacking strategy, the proposed approach aims to enable scalable and noise-tolerant estimation of modal parameters via FRVF using \textit{non-square} MIMO data, facilitating its deployment in realistic aircraft testing scenarios.

According to the main aim, the objectives of this work are as follows:
\begin{enumerate}
\item[i.] Generalise the FRVF formulation to non-square MIMO systems using an enhanced input-stacking approach;
\item[ii.] Assess the accuracy and noise robustness of the method through numerical simulations on a beam model;
\item[iii.] Evaluate its performance for applications to experimental aircraft GVT data compared with established SI techniques and literature results.
\end{enumerate}

\section{METHODOLOGY}

Classical SI techniques for modal analysis seek to determine the structure's $f_n$, $\zeta_n$, and $\mathbf{\phi}_n$ from experimentally measured FRFs. Within this class of methods, FRVF has recently been proposed as an alternative to traditional methods due to its solid frequency-domain formulation and its capability to represent broadband dynamic behaviour with high robustness \cite{Civera2021}.

\subsection{Fast and Relaxed Vector Fitting}

The original Vector Fitting (VF) algorithm approximates the measured FRFs using rational functions through an iterative process in which a predefined set of poles is successively relocated until convergence is achieved \cite{Gustavsen1999a}. The resulting model is expressed, in the $s$-domain, as
\begin{equation}
H(s) = \sum_{n=1}^{N} \frac{c_n}{s - a_n} + d + se
\label{eq:FRVF}
\end{equation}
where $a_n$ and $c_n$ denote the poles of the system and their corresponding residues, and $d$ and $e$ represent constant and proportional terms, respectively.

Importantly and interestingly, this formulation, which was derived, as said, in a field unrelated to structural dynamics \cite{Gustavsen1999a}, mirrors the pole-residue form of Single-Input Single-Output (SISO) FRFs in a multi-degree-of-freedom system, with (single) output index $i$ and (single) input index $j$:
\begin{equation}
H_{ij}(s) =\frac{X_i(s)}{F_j(s)} =\sum_{n=1}^{N} (\frac{c_{n}^{(ij)}}{s - a_n} + \frac{c_{n}^{(ij)*}}{s - a_n^*}) + d_{ij} + se_{ij}
\label{eq:MDoF FRF}
\end{equation}

Where $*$ denotes complex conjugate and $d_{ij} + se_{ij}$ only serves to accommodate the effects of out-of-band modes, below the lower cutoff frequency and above the upper one. Recall also that complex-conjugate poles and residues always refer to the same mode. As is widely known, the modal parameters can be obtained directly from the pole locations $a_n$, estimated from Equation~\eqref{eq:FRVF} (if SISO) or Equation~\eqref{eq:MDoF FRF} (MIMO), as:

\begin{equation}
f_n = \frac{ \sqrt{ \text{Re}(a_n)^2 + \text{Im}(a_n)^2 }}{ 2 \pi }\quad,
\end{equation}
\begin{equation}
\zeta_n = - \frac{ \text{Re}(a_n) }{ \sqrt{ \text{Re}(a_n)^2 + \text{Im}(a_n)^2 } }
\end{equation}

while the residues $c_n$ encode the modal participation factors and thus contain the mode shape information for each output channel. 

Also note that Equation~\eqref{eq:MDoF FRF} represents a receptance FRF (mN\textsuperscript{-1}), with displacements in output, but using velocities (mobility FRF in ms\textsuperscript{-1}N\textsuperscript{-1}) or accelerations (inertance FRF in ms\textsuperscript{-2}N\textsuperscript{-1}) would not change the poles, while the residues would only scale with the power of the pole, according to the number of time derivatives applied to the output, thus changing their amplitude but not their proportions (which is what matters in mode shapes).

Subsequently, the Relaxed Vector Fitting (RVF) introduces a pole-relaxation mechanism that removes the strict constraint of preserving the initial pole grid during the relocation process \cite{Vector2006}. This modification significantly improves convergence behaviour and reduces sensitivity to the initial pole selection, particularly in the presence of measurement noise. Nevertheless, RVF alone does not fully address the computational burden associated with large-scale problems.

From there, FRVF builds upon RVF by incorporating several algorithmic enhancements \cite{Deschrijver2008}:
\begin{itemize}
\item \textbf{QR-based residue estimation}, which improves numerical robustness and reduces computational and memory requirements for large MIMO datasets;
\item \textbf{Weak inverse-magnitude weighting}, which emphasises low-amplitude regions of the FRFs without excessively amplifying noise;
\item \textbf{Efficient pole stabilisation}, allowing convergence of poles and residues within a limited number of iterations.
\end{itemize}
As a result, FRVF enables accurate estimation of modal parameters even under high noise levels, while typically requiring lower model orders than classical VF formulations, as shown in \cite{Civera2021,Civera2021a}.

The description above summarises the essential features of the standard FRVF algorithm, which serves as the baseline for the developments presented in this work. In the recent literature, other additional refinements and alternative formulations have been proposed, including parallel implementations of VF \cite{Chinea2011}, pole-adding and skimming strategies \cite{Grivet-Talocia2006}, and time-domain extensions \cite{Grivet-Talocia2003}. These are not considered in this work.

\subsection{Enhanced Input Stacking}

The enhanced formulation proposed in this work builds upon the MIMO extension of VF originally introduced in \cite{Gustavsen2003}\footnote{As of 1 April 2026, a MATLAB implementation is available at \url{https://www.sintef.no/en/software/vector-fitting/downloads/matrix-fitting-toolbox/}.}. The classical MIMO FRVF framework was developed under the assumption of \textit{square} FRF matrices, in which the number of inputs equals the number of measured outputs. This assumption is well suited to electrical network modelling, where individual ports can be readily excited and measured interchangeably, making roving input–roving output testing straightforward and square FRF matrices the norm.

However, in the context of mechanical and structural testing, this assumption is rarely satisfied. Input and output channels are inherently different, particularly for large-scale structures, where a limited number of electrodynamic or hydraulic actuators is typically used in combination with a much larger number of sensors. As a result, GVTs almost invariably yield non-square FRF matrices, usually with the number of outputs exceeding the number of inputs by an order of magnitude. This mismatch constitutes a fundamental limitation for the direct application of the classical MIMO FRVF formulation.

To address this issue, an \textit{enhanced input stacking} strategy is introduced, enabling FRVF to efficiently process arbitrary \textit{non-square} MIMO configurations and thereby extending its applicability to realistic aircraft GVT scenarios. The proposed stacking procedure can be summarised as follows:

\begin{enumerate}
\item[i.] The internal loop dimensions of the algorithm are reformulated to accommodate non-square MIMO systems;
\item[ii.] The original three-dimensional FRF tensor, with dimensions $[N_{\text{out}} \times N_{\text{in}} \times N_{\text{freq}}]$, is reshaped into a two-dimensional array compatible with vector fitting;
\item[iii.] Input channel superposition stacking is performed frequency-wise to generate \textit{amplified} FRFs for each output channel.
\end{enumerate}

This last step is described in Algorithm 1.
\begin{algorithm}[!t]
\caption{Proposed full stacking procedure for non-square MIMO FRFs with input-channel superposition}
\label{alg:frf_stacking}
\begin{algorithmic}[1]
\State \textbf{Input:} $\mathrm{FRF}\in\mathbb{C}^{N_{\text{out}}\times N_{\text{in}}\times N_{\text{freq}}}$
\State \textbf{Output:} $f\in\mathbb{C}^{N_{\text{out}}\times N_{\text{freq}}}$
\State $f \gets \mathbf{0}_{N_{\text{out}}\times N_{\text{freq}}}$
\For{$row = 1$ to $N_{\text{out}}$}
    \For{$col = 1$ to $N_{\text{in}}$}
        \State $f(row,:) \gets f(row,:) + \mathrm{squeeze}\!\left(\mathrm{FRF}(row,col,:)\right)^{T}$
    \EndFor
\EndFor
\end{algorithmic}
\end{algorithm}

As can be seen, stacking is done per frequency bin and FRFs from different inputs are summed over the outputs, enforcing consistent alignment across inputs.

Algorithm 1 ensures that the non-square MIMO FRF is properly reshaped into a usable format for the FRVF algorithm, meeting the input requirements (a 2D array) while preserving the data structure. 

Notably, if the number of input and output channels coincides (i.e., $N_{in}=N_{out}=N_c$) and superposition is omitted, Algorithm 1 reduces to a simple full-matrix stacking, for which the conventional form suggested for square MIMO FRF \cite{Deschrijver2008} (Algorithm 2) is a restricted (upper-triangular) subset. Algorithm 2 is correct if reciprocity is assumed; however, while reciprocal square MIMO systems are conventional in multiport modelling for electronic engineering, the requirement of collocated actuators/sensors is too restrictive and impractical in modal testing for mechanical systems, for the reasons expressed previously.

In the proposed Algorithm 1, since the system is supposedly linear time-invariant, the principle of modal superposition holds. According to that, the sum of FRFs between inputs is equal to an effective FRF between output $i$ and a virtual input that excites all physical inputs with equal unit amplitude and phase; i.e., for any output channel, the corresponding estimated $\tilde{H}_i(s)$ represents a virtual, equivalent SISO system, driven by a distributed, coherent excitation, with equal weighting and phase across the inputs:

\begin{equation}
\tilde{H}_i(s) =  \sum_{j=1}^{J} H_{ij}(s) 
\end{equation}

Where $J$ is the total number of input channels considered.
As long as the system is linear (or weakly nonlinear, accepting a small error), the summation does not affect the common poles since they are invariant under linear combinations of system responses (i.e., transfer functions).
Furthermore, the summation actually increases modal participation, boosting the observability of weakly excited modes (i.e., with low residues at some input-output combinations), thus improving the signal-to-noise ratio (SNR) of poorly observable modes.
Although residues are therefore reweighted accordingly, this is inconsequential for mode shape estimation. In fact, mode shapes are affected in amplitude but not in spatial pattern; being all modal coefficients multiplied similarly, their proportions remain unaltered (that is to say, $\phi_n \propto \tilde{\phi}_n =\alpha_n \phi_n $ if $\alpha_n \in \mathbb{C}\setminus \{0\} $). Given that scaling is arbitrary anyway, any standard normalisation (max-norm, mass-norm, reference-DoF) will automatically remove $\alpha_n$, since
$\frac{\tilde{\phi}_n}{\lVert \tilde{\phi}_n \rVert} =  \frac{\phi_n}{\lVert \phi_n \rVert}$.

\begin{algorithm}[!t]
\caption{Conventional procedure (upper-triangular stacking of square MIMO FRF without input superposition)}
\label{alg:frf_stacking_triangular}
\begin{algorithmic}[1]
\State $tell \gets 0$
\For{$col = 1$ to $N_c$}
    \For{$row = col$ to $N_c$}
        \State $tell \gets tell + 1$
        \State $f(tell,:) \gets \text{squeeze}\!\left(\text{FRF}(row,col,:)\right)^{T}$
    \EndFor
\EndFor
\end{algorithmic}
\end{algorithm}

Hence, by aggregating multiple excitation contributions, each stacked output FRF embeds richer dynamic content, leading to an improved SNR while preserving the physical correlation between inputs and outputs. This structure ensures that the poles identified from the stacked representation remain unchanged, allowing direct estimates of natural frequencies and damping ratios, while the residues can be consistently recombined to recover physically meaningful mode shapes $\mathbf{\phi}_n$.

\subsection{Modal Parameters Identification}
Following the stacking procedure, the resulting FRFs are processed using FRVF. Then, the modal parameters are extracted through stabilisation diagrams to enforce physical validity. Candidate poles are retained here on the basis of three complementary stability criteria:

\begin{itemize}
\item \textbf{Frequency stability}, which requires the variation between successive model orders to remain below 1~Hz;
\item \textbf{Damping consistency}, rejecting poles that exhibit damping variations greater than 0.05;
\item \textbf{Mode shape correlation}, assessed through the Modal Assurance Criterion (MAC), which must exceed 0.95 between consecutive estimates.
\end{itemize}

With respect to algorithmic meta-parameters, FRVF requires the specification of an initial pole set, defined as a collection of complex-conjugate pole pairs uniformly distributed over the frequency range of interest to ensure adequate coverage of all potential structural resonances \cite{Gustavsen1999a}. To enhance numerical conditioning and mitigate the influence of measurement noise, a weak inverse-magnitude weighting scheme is employed during the fitting process, given by
\begin{equation}
weight = \frac{1}{\sqrt{|f(s)|}}
\end{equation}
where $f(s)$ is the target, experimentally determined FRF, which is to be approximated by the analytical transfer function $H(s)$. The FRVF procedure is subsequently executed for five iterations, which has been experimentally found to provide consistent convergence for pole locations in most applications \cite{Civera2021}.

\section{NUMERICAL CASE STUDY}
\begin{figure}[htp!]
    \centering
        \includegraphics[width=\columnwidth]{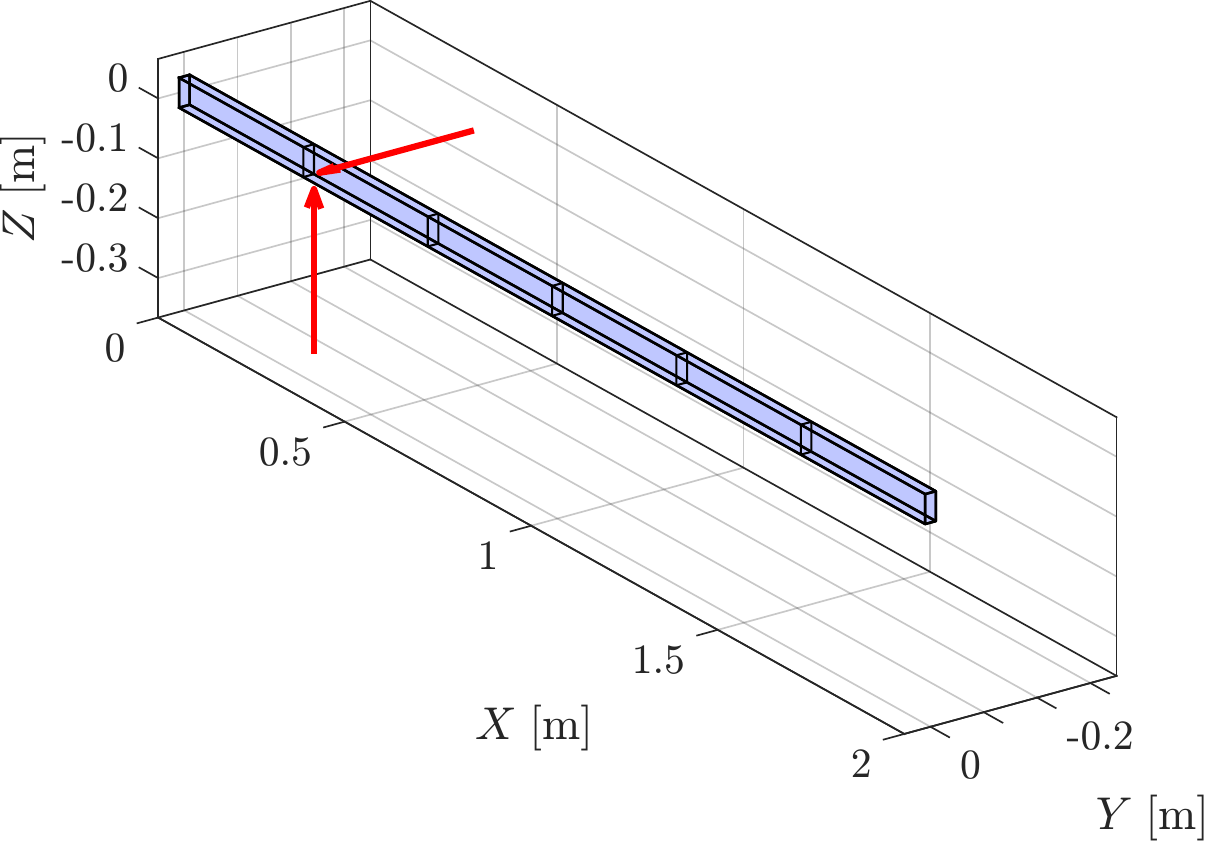}
        \caption{MIMO beam model with two excitation forces.} 
        \label{fig:3d_repre}
\end{figure}

The numerical model employed to preliminarily validate the proposed implementation of MIMO FRVF is based on a cantilever beam model (\cref{fig:3d_repre}), clamped at the root, corresponding to node 0. The beam has a total span of $L =$ 2 m and is characterised by a rectangular hollow cross-section. The external dimensions are $b_{\mathrm{ext}} =$ 0.02 m and $h_{\mathrm{ext}} =$ 0.05 m, with a uniform wall thickness of $t =$ 0.005 m.

The structure is assumed to be made of aluminium, with mass density $\rho =$ 2700 kgm\textsuperscript{-3}, Young's modulus $E =$ 69 {GPa}, and Poisson's ratio $\nu =$ 0.3, from which the shear modulus $G$ is derived. The beam is discretised into $N_{\mathrm{elem}} =$ 6 three-dimensional beam elements. This results in 6 nodes (excluding the fixed constraint) with 2 degrees of freedom (DoFs) each, along the $z-$ and $y-$ directions, totalling 12 output channels.
The mass and stiffness matrices $\mathbf{M}$ and $\mathbf{K}$ are assembled using the classic Euler--Bernoulli formulation extended to account for bending in two orthogonal planes.

Modal properties are obtained through the generalised eigenvalue problem associated with $\mathbf{K}$ and $\mathbf{M}$. Structural damping is introduced under the assumption of uncoupled modal damping, prescribing a constant damping ratio of $3\%$ for all modes. The resulting diagonal modal damping matrix is subsequently projected back into the physical coordinates to construct the global damping matrix $\mathbf{C}$.

The excitation configuration consists of two simultaneous unit impulse forces applied in orthogonal directions ($z$ and $y$) at the first free node of the beam, as illustrated in \cref{fig:3d_repre}. Each impulse has an amplitude of 1 N and is applied at $t =$ 0 s for a duration equal to one sampling interval. The system response is simulated using a sampling frequency of $f_s =$ 1000 {Hz} over a total duration of $T_{total} =$ 30 s, ensuring adequate frequency resolution and compliance with the Nyquist criterion for the first 10 modes of interest.

\begin{figure}[htp!] 
  \centering
    \includegraphics[width=\columnwidth]{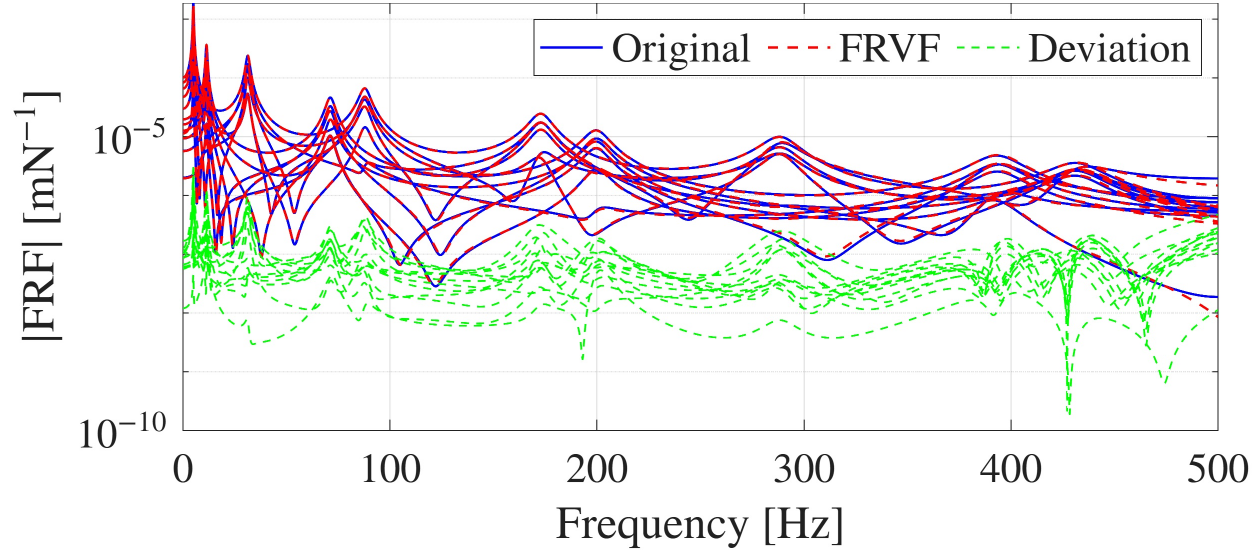}
  \caption{Magnitude of the FRF data, fit, and deviation for the noise-free MIMO beam.}
  \label{fig:3d_comp}
\end{figure}

The resulting displacement ($z$- and $y$-directions) responses are extracted at multiple degrees of freedom along the span, forming a MIMO dataset used for subsequent frequency response function estimation and modal identification. This numerical setup provides a controlled yet sufficiently rich dynamic environment to assess the performance of the proposed identification framework.
The system response is shown in \cref{fig:3d_comp}, with the magnitude of the FRF. 

It can be observed that the 10 dominant modes of interest are included within the frequency range observable with the selected sampling frequency. Moreover, the FRVF fitting accurately matches the original FRF response. This is shown by the deviation, expressed in terms of the Root Mean Square Error (RMSE), which is orders of magnitude lower than the original FRF amplitude.

\Cref{tab:3d_beam_modes} reports the modal parameters identified by the proposed FRVF MIMO method, the benchmark method (LSCE), and the analytical results obtained from eigenanalysis of the system matrices. It is worth noting that the $\mathbf{\phi}_n$ comparison is presented using MAC values, with the analytical results used as reference. 

\begin{table*}[!htb]
\renewcommand{\arraystretch}{1.25}
\caption{Comparison of analytical, FRVF, and LSCE estimates of natural frequencies, damping ratios, and MAC values for the MIMO beam.}
\label{tab:3d_beam_modes}
\centering
\small
\begin{tabular}{|c|ccc|ccc|cc|}
\hline
\bfseries Mode
& \multicolumn{3}{c|}{\bfseries Natural Frequency [Hz]}
& \multicolumn{3}{c|}{\bfseries Damping Ratio [\textendash]}
& \multicolumn{2}{c|}{\bfseries MAC} \\
\hline
& \bfseries Analytical & \bfseries FRVF & \bfseries LSCE
& \bfseries Analytical & \bfseries FRVF & \bfseries LSCE
& \bfseries FRVF & \bfseries LSCE \\
\hline\hline
1  & 5.001   & 5.001   & 5.001   & 0.03 & 0.03 & 0.03 & 1  & 1  \\ \hline
2  & 11.367  & 11.367  & 11.367  & 0.03 & 0.03 & 0.03 & 1  & 1  \\ \hline
3  & 31.347  & 31.347  & 31.347  & 0.03 & 0.03 & 0.03 & 1  & 1  \\ \hline
4  & 71.253  & 71.253  & 71.253  & 0.03 & 0.03 & 0.03 & 1  & 1  \\ \hline
5  & 87.912  & 87.912  & 87.912  & 0.03 & 0.03 & 0.03 & 1  & 1  \\ \hline
6  & 173.065 & 173.065 & 173.065 & 0.03 & 0.03 & 0.03 & 1  & 1  \\ \hline
7  & 199.826 & 199.826 & 199.826 & 0.03 & 0.03 & 0.03 & 1  & 1  \\ \hline
8  & 288.530 & 288.530 & 288.530 & 0.03 & 0.03 & 0.03 & 1  & 1  \\ \hline
9  & 393.383 & 393.383 & 393.383 & 0.03 & 0.03 & 0.03 & 1  & 1  \\ \hline
10 & 431.710 & 431.710 & 431.710 & 0.03 & 0.03 & 0.03 & 1  & 1  \\ \hline
\end{tabular}
\end{table*}

As shown, all existing modes are identified by FRVF and LSCE. The associated $f_n$ errors between the identified values (FRVF and LSCE) and the analytical results are negligible ($<$ 0.1\%), and the same is true for the identified  $\zeta_n$. A similar behaviour is found for $\mathbf{\phi}_n$, where all MAC values are 1. This shows an almost perfect match between the expected and identified $\mathbf{\phi}_n$. LSCE performed better than expected since it has been shown, in previous works \cite{Dessena2024}, to struggle in a MIMO system with some closely-spaced modes. Nevertheless, the frequency range investigated here is smaller than that taken into account in the aforementioned work. An additional basis for the comparison between LSCE and FRVF will be offered in the experimental case study (Section~\ref{sec:bae-hawk-t1a}).

It is worth noting that the results obtained are derived from stabilisation diagrams, such that numerically stable poles can be extracted from higher model orders value. 

\subsection{Robustness to noise}
To quantify the sensitivity of the proposed approach to measurement noise, a systematic noise analysis is conducted. Artificial noise is progressively introduced into both input and output signals simultaneously, allowing the distinct effects of each noise source on the identified system parameters to be assessed. To keep this discussion brief, only input-output noise is considered here, as it has been shown to be the most representative case in this type of study \cite{Dessena2022g}. Noise is modelled as Additive White Gaussian Noise (AWGN), whose level is defined as a percentage of the standard deviation of the response signal. For this parametric noise analysis, the following levels are considered: 0 (noise-free baseline), 0.1, 0.3, 0.5, 0.7, 1, 3, and 5\%.

Preliminarily, it is worth noting that any modal parameters presented in this section are obtained from the inspection of stabilisation diagrams, the clarity and univocity of which are negatively affected by noise (see, e.g., \cite{Civera2021,Civera2021a}), as the quality of the fitting decreases. Nevertheless, in these cases, this will allow for the identification of all modes, although with varying accuracy, related to the corresponding noise level. Furthermore, only the noise robustness of MIMO FRVF is assessed here since FRVF-retrieved modal parameters are already found to be in great agreement with LSCE and the analytical values for the noiseless case. 

With the aim of assessing the MIMO FRVF robustness to noise, the modal parameters identified in the noise cases are compared to the analytical (noise-free) counterpart. These results are shown in \cref{fig:3d_noise_combined} as relative differences, reported in percentages, for $f_n$ and $\zeta_n$ and as MAC values (0 no correlation, 1 perfect correlation) with respect to analytical values for $\mathbf{\phi}_n$.

\begin{figure}[htp!]
    \centering
    \begin{subfigure}[t]{\columnwidth}
        \centering
        \includegraphics[width=\columnwidth]{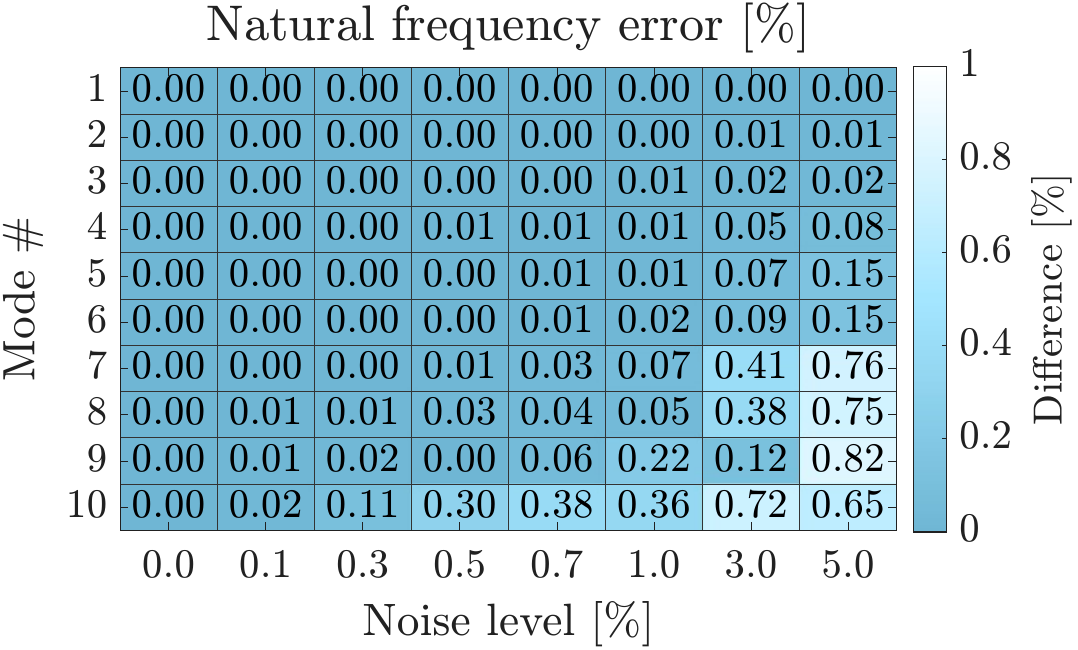}
        \subcaption{}
        \label{fig:3d_noise_freq}
    \end{subfigure}
    \begin{subfigure}[t]{\columnwidth}
        \centering
        \includegraphics[width=\columnwidth]{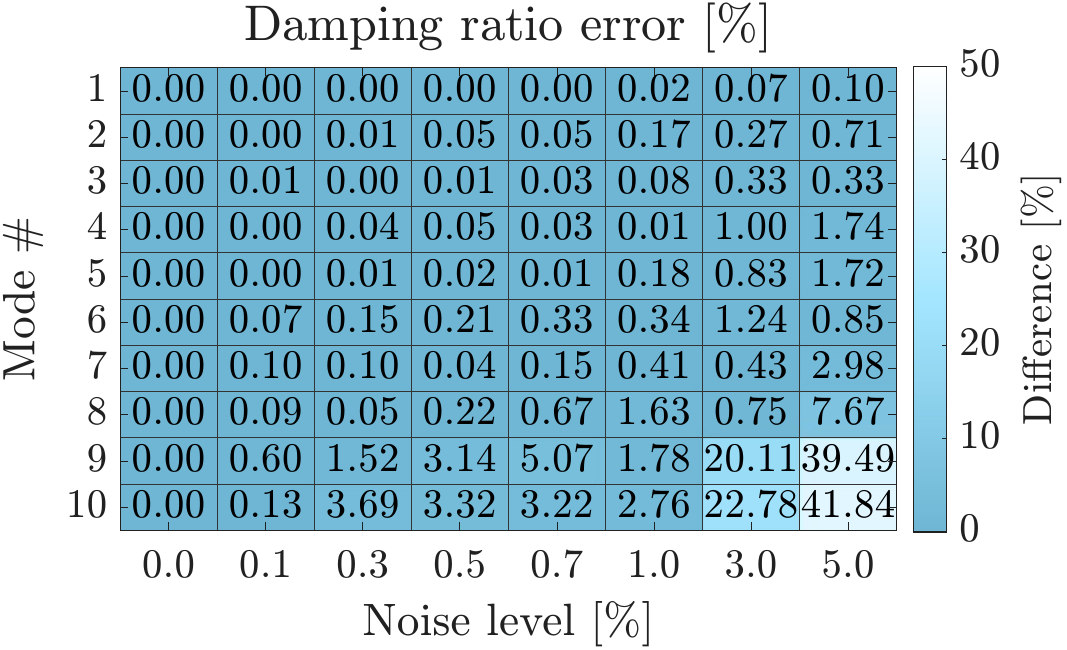}
        \subcaption{}
        \label{fig:3d_noise_damp}
    \end{subfigure}
\end{figure}
\begin{figure}[htp!]\ContinuedFloat
    \centering
    \begin{subfigure}[t]{\columnwidth}
        \centering
        \includegraphics[width=\columnwidth]{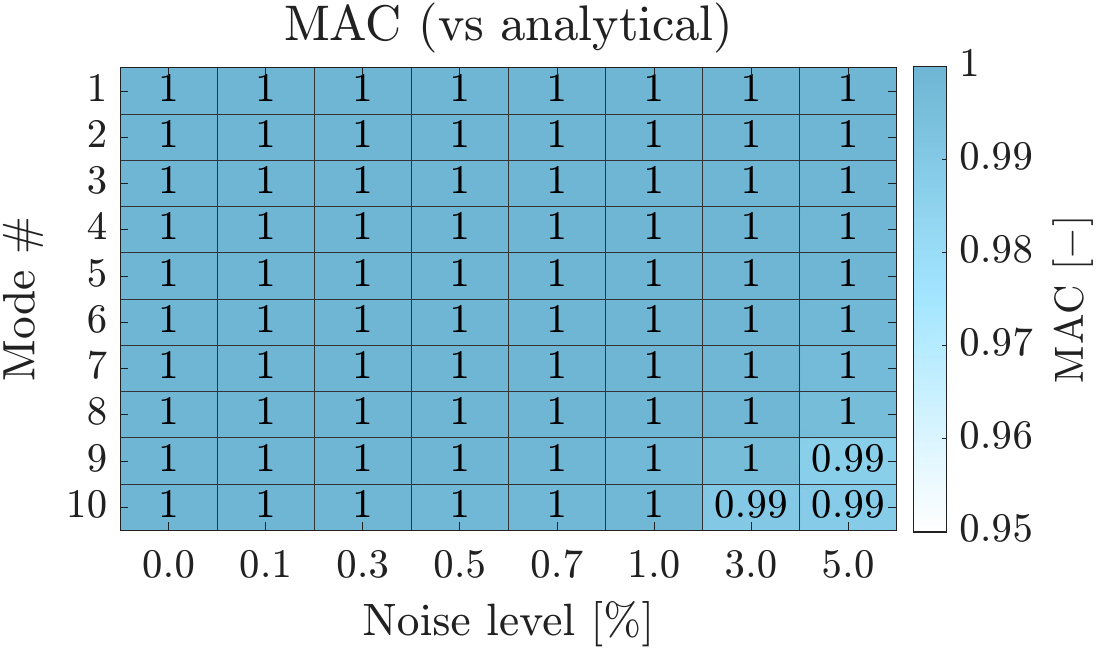}
        \subcaption{}
        \label{fig:3d_noise_mac}
    \end{subfigure}
    \caption{Sensitivity analysis of the MIMO beam under increasing noise levels: (a) frequency error, (b) damping ratio error, and (c) MAC values with respect to the analytical values, for the combined input--output noise case.}
    \label{fig:3d_noise_combined}
\end{figure}

The robustness of FRVF in a MIMO setting is illustrated in \autoref{fig:3d_noise_combined}. Although the identified natural frequencies and damping ratios remain mostly stable across noise levels up to 1\%, the presence of input-output noise exceeding 3\% -- particularly for higher modes, 9 and 10 -- leads to a decrease in the identification quality. In this regime, the MAC values and the identified $f_n$ still show an almost perfect correlation with the analytical values. The former shows MAC values of 0.99 and maximum frequency deviations of less than 0.9\%. The same is not shown for $\zeta_n$, as the relative error starts to be noticeable, reaching $\geq$20\% and $\geq$40\% for the higher modes at 3 and 5\% noise level.

These results show that the proposed MIMO FRVF approach demonstrates robust performance across the entire configuration, successfully identifying nearly all vibration modes with frequency errors below 1\% and MAC values $\geq$ 0.99. Even under more adverse noise conditions, the method remains effective in capturing the dominant FRF peaks, while only the damping ratios $\zeta_n$ of the higher modes are not appropriately captured. It is important to note that, despite the noise, all modes are always retrieved (even if in a degraded form).

In closing the numerical case study, it can be asserted that, at least in the investigated scenario, MIMO FRVF offers a precise and robust, noise-resistant, modal identification parameter capability, as compared to analytical and LSCE-derived results. This will be also confirmed on experimental data.

\section{BAE SYSTEMS HAWK T1A TRAINER JET AIRCRAFT} 
\label{sec:bae-hawk-t1a}
\begin{figure}[htp!]
     \centering
     \includegraphics[width=\linewidth]{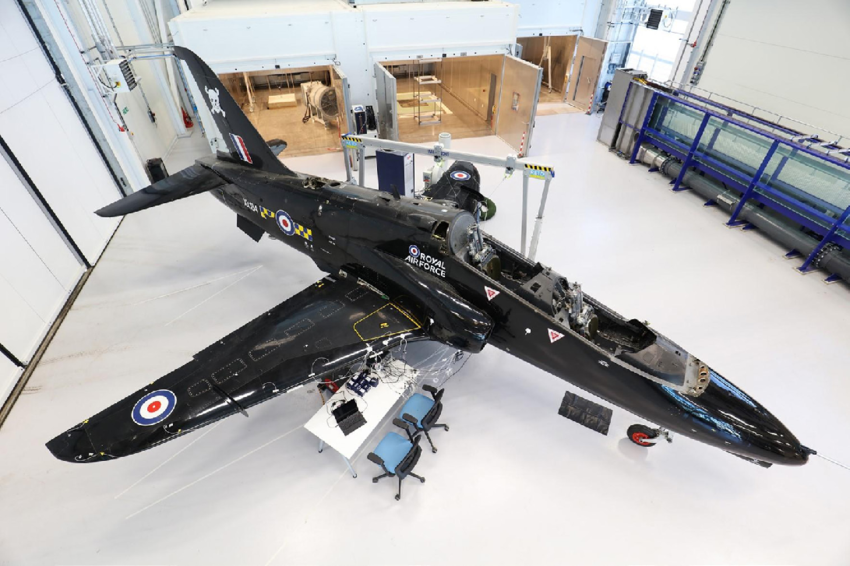}
     \caption{The BAE Systems Hawk T1A trainer jet aircraft at the LVV (Sheffield, UK -- retrieved from \cite{Wilson2024}).}
     \label{fig:hwk}
 \end{figure}

This experimental case study considers a BAE Systems Hawk T1A trainer jet aircraft (\autoref{fig:hwk}) \cite{Wilson2024}, commonly referred to as the Hawk. The aircraft frame was donated to the University of Sheffield by the Defence Science and Technology Laboratory of the Royal Air Force (RAF, UK) and is currently housed in the Laboratory for Verification and Validation (LVV) of the University. The aircraft previously served the RAF in advanced pilot training and has since been decommissioned. Although the structure is largely intact, several items have been removed, including the engine, cockpit canopy, selected wing flaps, avionics, and auxiliary systems. The aircraft rests on its landing gear in a stationary setup on the facility floor, with the tyres intentionally deflated to reduce support stiffness and approximate free-free conditions as closely as practicable without suspension.

The test setup employs five excitation sources, carefully positioned to cover the primary structural components. Specifically, one exciter is mounted on each wing (port and starboard), one on the vertical stabiliser, and one on each side of the horizontal stabiliser. To monitor the vibration response, a total of 85 accelerometers are deployed across the aircraft frame, comprising uniaxial and triaxial sensors, which provide 91 response channels in total. In total,
216 tests are available in the dataset\footnote{\url{https://orda.shef.ac.uk/articles/dataset/BAE_T1A_Hawk_Full_Structure_Modal_Test/24948549}} . For the scope of this work, the ten repetitions of the highest level (0.5 V) white noise input test repetitions are considered\footnote{This set is labelled in \cite{Wilson2024} as HS\_WN/5.}.

The measured FRFs originally span a broad frequency range, extending up to 1024~Hz (sampling frequency $f_s=$ 2048 Hz). However, for the objectives of this study, such an extensive range is not required. The primary modes of interest (namely, the global structural modes, with the highest portion of the participating mass) occur at lower frequencies, allowing the upper bound to be reduced without losing relevant information.

Consequently, the FRFs are truncated to the 5–165~Hz range. This specific frequency band is also chosen to enable a direct comparison with the results reported in \cite{Dessena2024, Dessena2025a}, obtained using the iLF, respectively for the full aircraft and for a subdataset of the starboard wing (described later). The selected range includes the first symmetric and antisymmetric wing bending modes, the primary fuselage bending mode, and the initial torsional modes, capturing the dynamic behaviour most relevant for flutter prediction and structural model updating.

After truncation, the FRF dataset is structured as follows:
\[
[91~\text{outputs} \times 5~\text{inputs} \times N_\text{freq}]
\]
where $N_\text{freq}$ represents the number of frequency samples within the 5–165~Hz interval. to reduce computational effort and processing time, while also increasing the signal-to-noise ratio, the ten available repetitions for each input–output pair were averaged prior to analysis using the proposed MIMO FRVF technique. 

\subsection{Starboard Wing Subdataset}
\begin{figure}
    \centering
    \includegraphics[width=\linewidth]{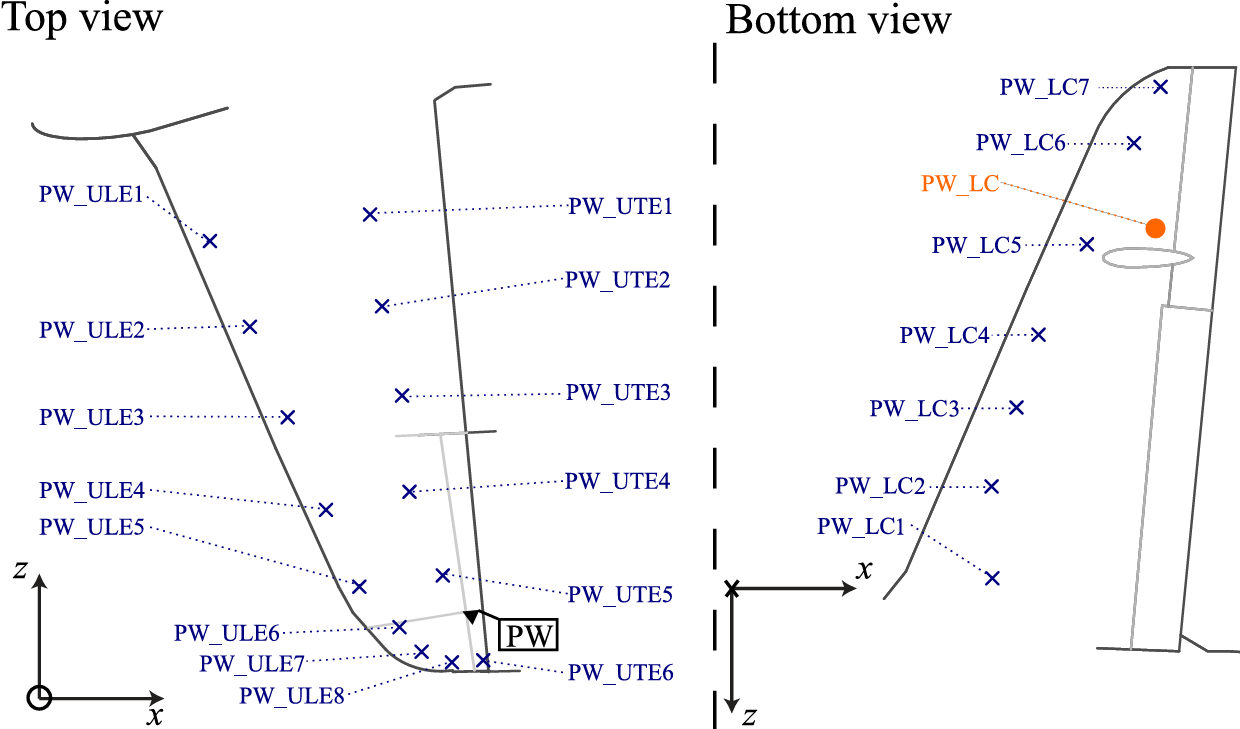}
    \caption{Schematic diagram showing the accelerometer locations on the wing (adapted from \cite{Wilson2024}).}
    \label{fig:wing}
\end{figure}

To allow a more focused analysis, the MIMO FRVF modal identification is first performed using only the 21 accelerometer channels on the starboard wing (see \autoref{fig:wing}). This configuration should not be conflated with the separate Hawk wing SIMO dataset reported in \cite{Haywood-Alexander2024}, as this configuration still considers the inputs from all five shakers within a MIMO framework. This set includes four shakers not mounted directly on the wing, which are therefore treated as \textit{remote} excitation sources, since they still contribute to the wing dynamic response.

In \autoref{tab:comparison_trunc_2}, the results are compared with the modes identified by iLF and LSCE in \cite{Dessena2025a}, both using the same sensor subset.

\begin{table}[!htp]
\caption{Comparison of modes identified by the proposed MIMO FRVF with MIMO iLF and LSCE ~\cite{Dessena2025a}, using the truncated dataset (5--165~Hz) for sensors placed on the starboard wing.}
\label{tab:comparison_trunc_2}
\centering
\small
\renewcommand{\arraystretch}{1.15}
\begin{tabular}{|c|ccc|ccc|}
\hline
\bfseries Mode
& \multicolumn{3}{c|}{\bfseries Natural Frequency [Hz]}
& \multicolumn{3}{c|}{\bfseries Damping Ratio [--]} \\
\hline
& \bfseries iLF & \bfseries FRVF & \bfseries LSCE
& \bfseries iLF & \bfseries FRVF & \bfseries LSCE \\
\hline\hline
1 & 7.03  & 7.055  & --     & 0.03 & 0.01 & --     \\ \hline
2 & 15.42 & 15.453 & 12.732 & 0.01 & 0.01 & 0.26 \\ \hline
3 & 16.28 & 16.327 & --     & 0.01 & 0.01 & --     \\ \hline
\end{tabular}
\end{table}

FRVF is clearly capable of detecting the three wing modes, closely matching the natural frequencies and the damping ratio values in \cite{Dessena2025a} (including the greater variability generally associated with estimates of $\zeta_n$). The extracted mode shapes are compared only for the full-aircraft case, as the mode shapes from the subdataset are much less informative. Nevertheless, it can be seen here how LSCE fails to properly capture all three modes in the starboard wing subdataset, since LSCE only identifies mode \#2 with a considerable deviation in $f_n$. 

\subsection{Full Aircraft Dataset}
\begin{figure}[htp!]
    \centering
    \includegraphics[width=\columnwidth]{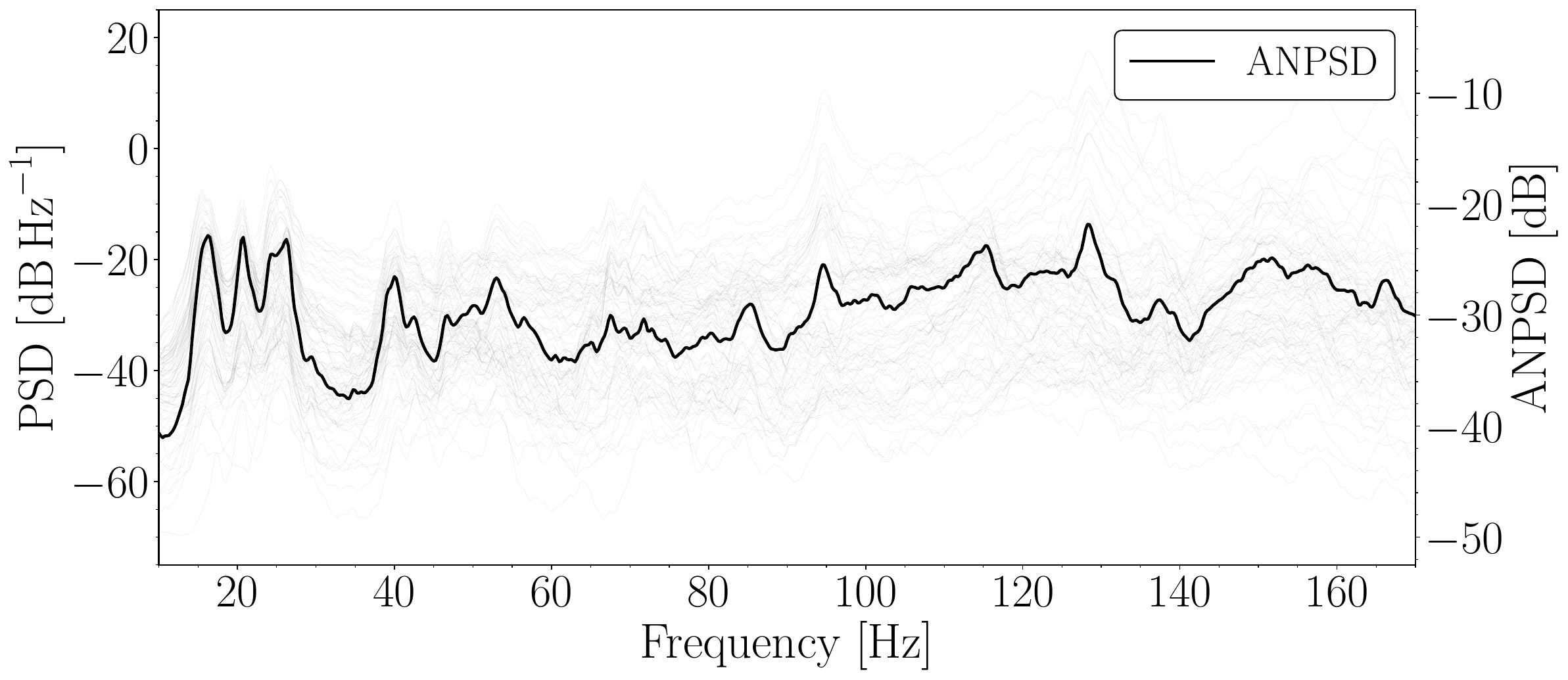}
    \caption{PSD representation across all 91 accelerometer channels, including the ANPSD, for the complete dataset}
    \label{fig:PSD}
\end{figure}

For the same reduced frequency range, the full aircraft dataset (91 output channels) is considered, with linear spacing of the initial poles and model order ranging from 30 to 80 for the MIMO FRVF identification. The range of model orders is based on the number of modes previously identified, 32. This estimation is made possible by accessing the results included in \cite{Dessena2024} and by inspecting the Average Normalised Power Spectral Density (ANPSD) of the system superimposed on the Power Spectral
Density (PSD) of all channels, as shown in \cref{fig:PSD}.

The complete set of identified natural frequencies and damping ratios 
is reported in \cref{tab:trunc_freq_damp}, whilst the corresponding MAC values are collected in 
\cref{tab:trunc_mac}. 

\begin{table*}[!htp]
\renewcommand{\arraystretch}{1.25}
\caption{Comparison of the natural frequencies and damping ratios as identified by iLF \cite{Dessena2024}, FRVF, and LSCE using the truncated dataset (5--165 Hz).}
\label{tab:trunc_freq_damp}
\centering
\small

\begin{minipage}[t]{0.49\textwidth}
\centering
\begin{tabular}{|c|ccc|ccc|}
\hline
\bfseries Mode
& \multicolumn{3}{c|}{\bfseries Natural Frequency [Hz]}
& \multicolumn{3}{c|}{\bfseries Damping Ratio [--]} \\
\hline
& \bfseries iLF & \bfseries FRVF & \bfseries LSCE
& \bfseries iLF & \bfseries FRVF & \bfseries LSCE \\
\hline\hline
1  & 6.976   & 7.019   & 9.110   & 0.028 & 0.010 & 0.668 \\ \hline
2  & 15.435  & 15.520  & 13.980  & 0.008 & 0.008 & 0.712 \\ \hline
3  & 16.319  & 16.285  & --      & 0.010 & 0.007 & --    \\ \hline
4  & 17.418  & --      & --      & 0.015 & --    & --    \\ \hline
5  & 20.679  & 20.650  & --      & 0.012 & 0.004 & --    \\ \hline
6  & 22.159  & --      & --      & 0.012 & --    & --    \\ \hline
7  & 24.199  & 24.260  & --      & 0.010 & 0.001 & --    \\ \hline
8  & 25.363  & 25.412  & --      & 0.013 & 0.013 & --    \\ \hline
9  & 26.294  & --      & --      & 0.016 & --    & --    \\ \hline
10 & 40.176  & 40.460  & --      & 0.021 & 0.012 & --    \\ \hline
11 & 43.292  & 42.786  & --      & 0.016 & 0.016 & --    \\ \hline
12 & 46.900  & 46.753  & --      & 0.017 & 0.009 & --    \\ \hline
13 & 49.999  & 50.408  & --      & 0.031 & 0.018 & --    \\ \hline
14 & 53.381  & 53.244  & --      & 0.024 & 0.026 & --    \\ \hline
15 & 67.590  & --      & --      & 0.006 & --    & --    \\ \hline
16 & 68.750  & 68.973  & 67.707  & 0.013 & 0.004 & 0.015 \\ \hline
\end{tabular}
\end{minipage}
\hfill
\begin{minipage}[t]{0.49\textwidth}
\centering
\begin{tabular}{|c|ccc|ccc|}
\hline
\bfseries Mode
& \multicolumn{3}{c|}{\bfseries Natural Frequency [Hz]}
& \multicolumn{3}{c|}{\bfseries Damping Ratio [--]} \\
\hline
& \bfseries iLF & \bfseries FRVF & \bfseries LSCE
& \bfseries iLF & \bfseries FRVF & \bfseries LSCE \\
\hline\hline
17 & 72.166  & 72.134  & --      & 0.019 & 0.015 & --    \\ \hline
18 & 81.018  & 81.585  & --      & 0.030 & 0.012 & --    \\ \hline
19 & 83.906  & 83.944  & --      & 0.016 & 0.007 & --    \\ \hline
20 & 94.701  & 94.361  & --      & 0.006 & 0.001 & --    \\ \hline
21 & 97.458  & 96.620  & --      & 0.025 & 0.016 & --    \\ \hline
22 & 102.146 & 102.831 & --      & 0.021 & 0.015 & --    \\ \hline
23 & 106.544 & 106.417 & --      & 0.017 & 0.015 & --    \\ \hline
24 & 115.050 & 115.445 & --      & 0.015 & 0.008 & --    \\ \hline
25 & 120.952 & 121.388 & --      & 0.015 & 0.023 & --    \\ \hline
26 & 125.377 & --      & --      & 0.013 & --    & --    \\ \hline
27 & 128.984 & 128.416 & --      & 0.015 & 0.001 & --    \\ \hline
28 & 132.213 & 132.230 & 132.847 & 0.007 & 0.008 & 0.002 \\ \hline
29 & 137.581 & 137.876 & --      & 0.009 & 0.005 & --    \\ \hline
30 & 149.674 & 149.946 & --      & 0.007 & 0.001 & --    \\ \hline
31 & 151.894 & 151.884 & --      & 0.009 & 0.001 & --    \\ \hline
32 & 156.916 & 157.048 & --      & 0.010 & 0.010 & --    \\ \hline
\end{tabular}
\end{minipage}

\end{table*}

\begin{table*}[!htp]
\renewcommand{\arraystretch}{1.25}
\caption{Comparison of FRVF and LSCE using the truncated dataset (5--165 Hz): MAC  values with respect to identifications by iLF \cite{Dessena2024}.}
\label{tab:trunc_mac}
\centering
\small

\begin{minipage}[t]{0.24\textwidth}
\centering
\begin{tabular}{|c|cc|}
\hline
\bfseries Mode & \multicolumn{2}{c|}{\bfseries MAC} \\
\hline
& \bfseries FRVF & \bfseries LSCE \\
\hline\hline
1 & 0.829 & 0.036 \\ \hline
2 & 0.895 & 0.023 \\ \hline
3 & 0.740 & --    \\ \hline
4 & --    & --    \\ \hline
5 & 0.918 & --    \\ \hline
6 & --    & --    \\ \hline
7 & 0.574 & --    \\ \hline
8 & 0.783 & --    \\ \hline
\end{tabular}
\end{minipage}\hfill
\begin{minipage}[t]{0.24\textwidth}
\centering
\begin{tabular}{|c|cc|}
\hline
\bfseries Mode & \multicolumn{2}{c|}{\bfseries MAC} \\
\hline
& \bfseries FRVF & \bfseries LSCE \\
\hline\hline
9  & --    & --    \\ \hline
10 & 0.069 & --    \\ \hline
11 & 0.007 & --    \\ \hline
12 & 0.521 & --    \\ \hline
13 & 0.290 & --    \\ \hline
14 & 0.415 & --    \\ \hline
15 & --    & --    \\ \hline
16 & 0.011 & 0.013 \\ \hline
\end{tabular}
\end{minipage}\hfill
\begin{minipage}[t]{0.24\textwidth}
\centering
\begin{tabular}{|c|cc|}
\hline
\bfseries Mode & \multicolumn{2}{c|}{\bfseries MAC} \\
\hline
& \bfseries FRVF & \bfseries LSCE \\
\hline\hline
17 & 0.737 & --    \\ \hline
18 & 0.198 & --    \\ \hline
19 & 0.059 & --    \\ \hline
20 & 0.889 & --    \\ \hline
21 & 0.062 & --    \\ \hline
22 & 0.244 & --    \\ \hline
23 & 0.026 & --    \\ \hline
24 & 0.010 & --    \\ \hline
\end{tabular}
\end{minipage}\hfill
\begin{minipage}[t]{0.24\textwidth}
\centering
\begin{tabular}{|c|cc|}
\hline
\bfseries Mode & \multicolumn{2}{c|}{\bfseries MAC} \\
\hline
& \bfseries FRVF & \bfseries LSCE \\
\hline\hline
25 & 0.763 & --    \\ \hline
26 & --    & --    \\ \hline
27 & 0.534 & --    \\ \hline
28 & 0.623 & 0.010 \\ \hline
29 & 0.571 & --    \\ \hline
30 & 0.501 & --    \\ \hline
31 & 0.051 & --    \\ \hline
32 & 0.850 & --    \\ \hline
\end{tabular}
\end{minipage}

\end{table*}

\begin{figure*}[!htp]
    \centering

    \begin{subfigure}[t]{0.32\linewidth}
        \centering
        \includegraphics[width=\textwidth]{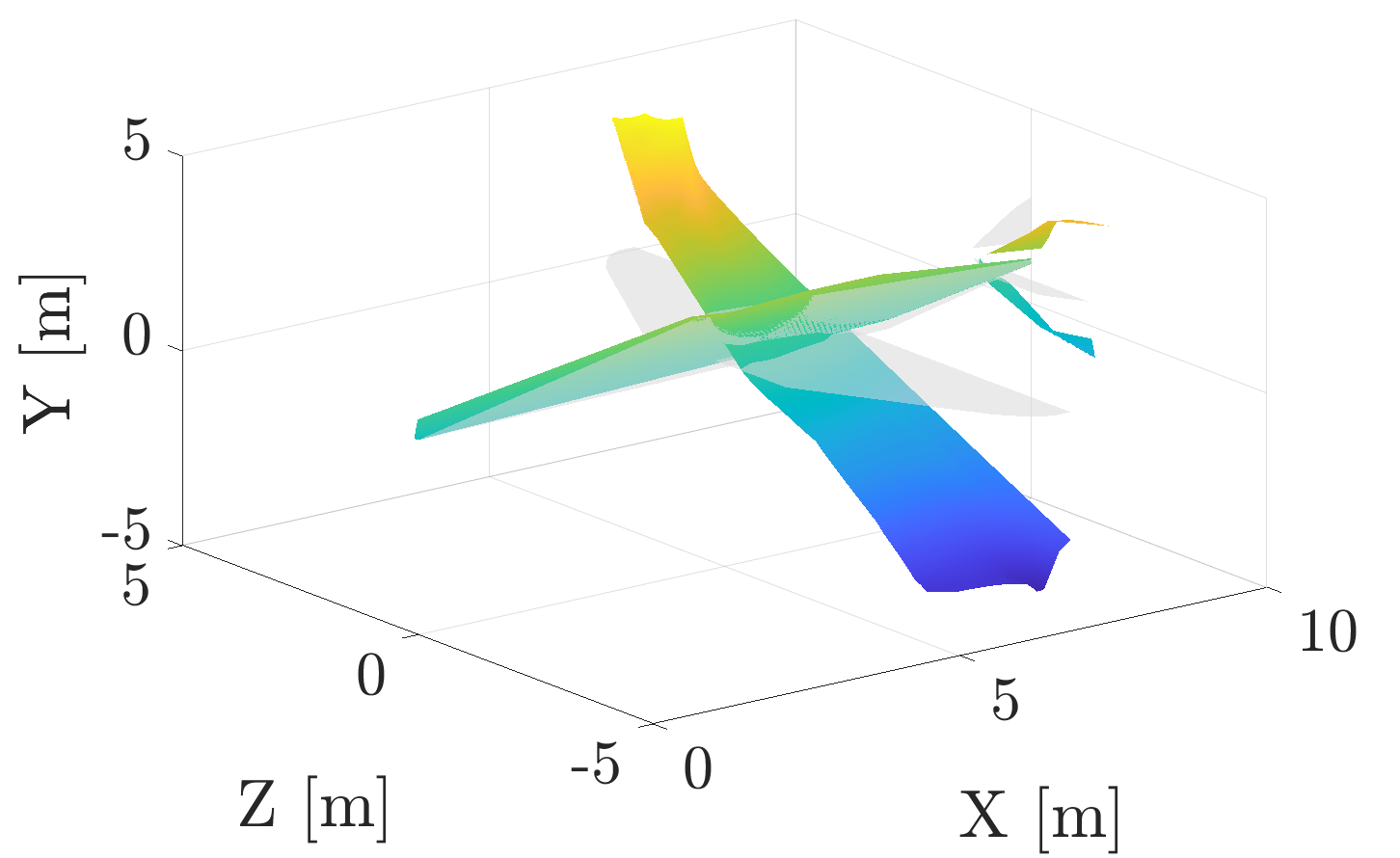}
        \subcaption{} 
        \label{fig:mode_1}
    \end{subfigure}
    \begin{subfigure}[t]{0.32\linewidth}
        \centering
        \includegraphics[width=\textwidth]{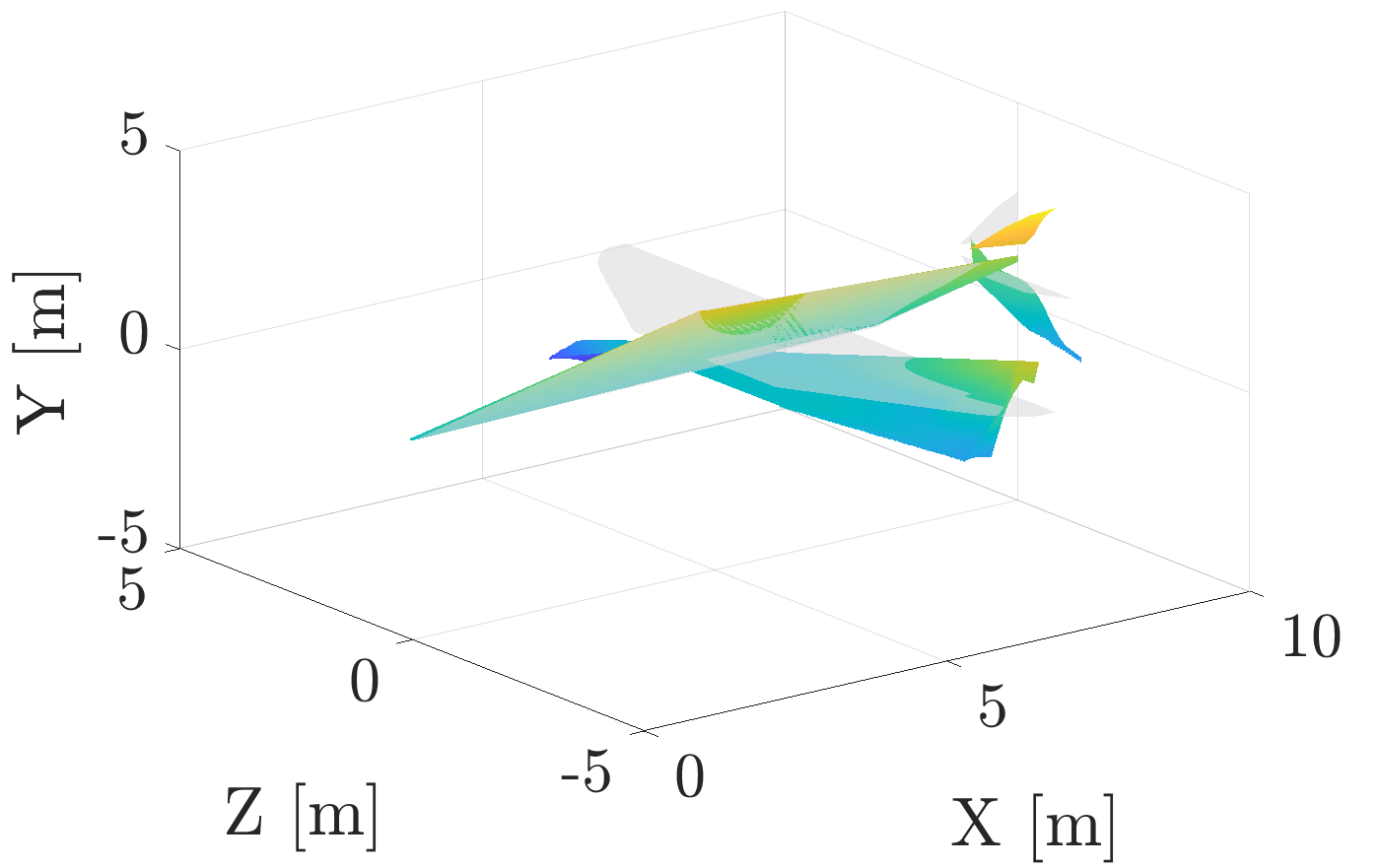}
        \subcaption{} 
        \label{fig:mode_2}
    \end{subfigure}
    \begin{subfigure}[t]{0.32\linewidth}
        \centering
        \includegraphics[width=\textwidth]{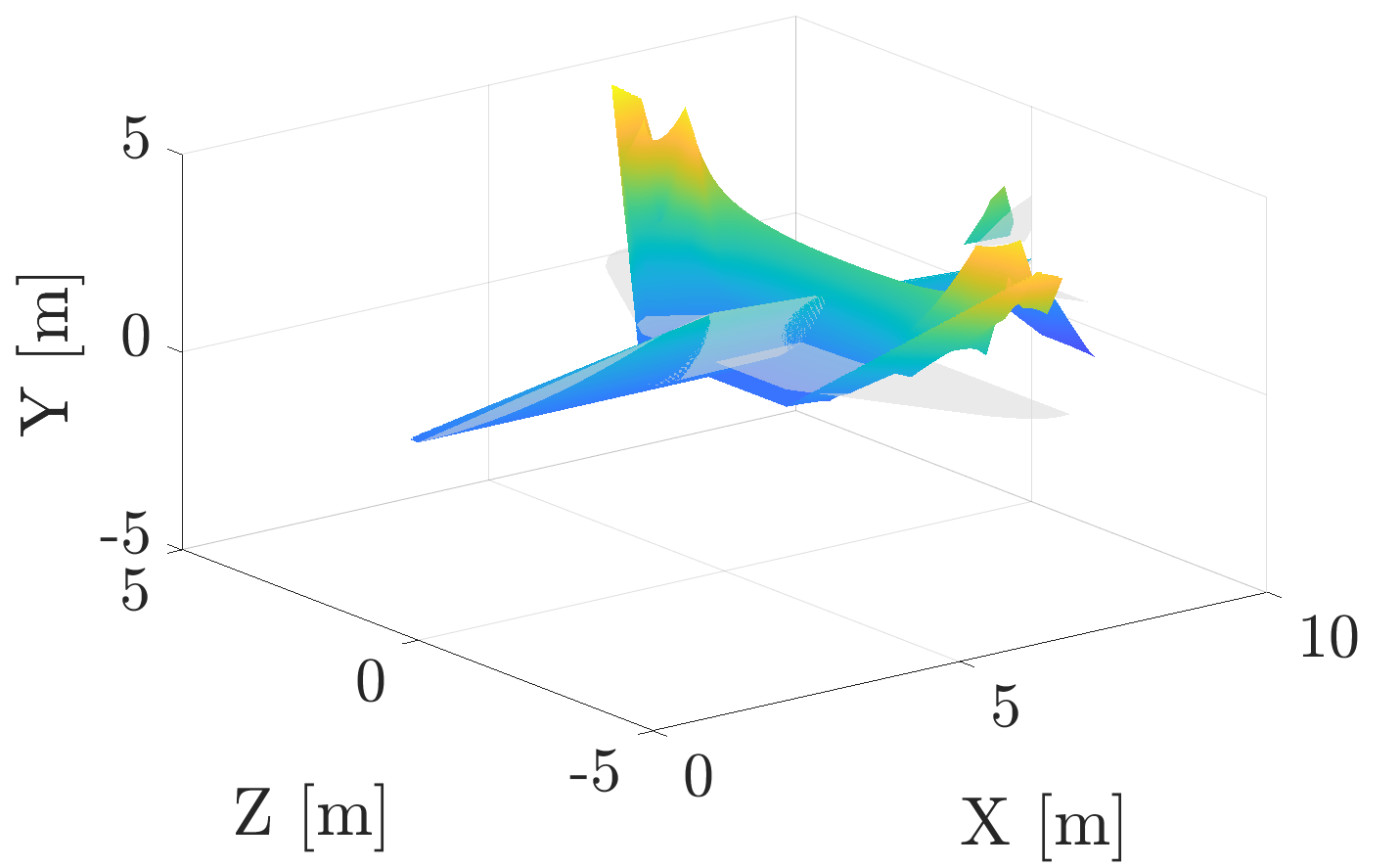}
        \subcaption{} 
        \label{fig:mode_3}
    \end{subfigure}

    \caption{First three mode shapes of the Hawk obtained using the full truncated dataset: (a) first mode, (b) second mode, and (c) third mode.}
    \label{fig:modes_hawk}
\end{figure*}

These three wing modes are represented in \autoref{fig:modes_hawk}.

From \cref{tab:trunc_freq_damp} and \cref{tab:trunc_mac} it is clear that,natural frequency-wise , FRVF successfully recovers 27 of the 32 modes identified by iLF, with $\Delta f_n$ generally within 1\% throughout the entire band. The largest deviation observed is approximately $|f_n,_{\text{FRVF}}-f_n,_{\text{iLF}}|/f_n,_{\text{iLF}}=$ 1.17\% in mode~11 (43.3~Hz). LSCE, by contrast, identifies only four modes --- and with substantial deviations in both $f_n$ (e.g., 30.5\% in mode~1) and $\zeta_n$ (e.g., 2285\% at mode~1) --- confirming its well-known limitations under broadband MIMO excitation~\cite{Dessena2024}. The progressive reduction in MAC values observed from mode~8 onwards (from 0.783 in mode~8 down to values below 0.1 for several modes above 40~Hz) is consistent with the increasing modal density and reduced SNR at higher frequencies, and with the inherently lower observability of modes that are either spatially localised or weakly excited by the available shaker configuration. Finally, the mode shapes shown in \cref{fig:modes_hawk} show very good visual agreement with those presented in \cite{Dessena2024}.

\section{DISCUSSION}

Across both validation scenarios, FRVF combined with enhanced input stacking demonstrates consistently strong performance. In the numerical beam case, all ten modes are recovered with negligible frequency errors ($<$0.1\%) and MAC values of unity under combined input--output noise levels up to 0.7\%. Damping ratios, whilst remaining accurate for modes~1--8 across the same range, exhibit progressive degradation for modes~9 and~10 even below 1\% noise, reaching 5.1\% and 3.2\% at 0.7\%, respectively. Appreciable degradation in both frequency and damping for these higher modes becomes evident only at noise levels of 3\% and above, where damping errors reach $\approx$20\% and $\approx$40\% respectively, whilst frequency errors remain below 0.9\% and MAC values remain $\geq$0.99.

For Hawk T1A, FRVF identifies 27 of the 32 modes reported by iLF within the 5--165~Hz band, with frequency agreement generally within 1\% and MAC values that are satisfactory for the lower, well-separated modes (0.829, 0.895, and 0.918 for modes~1, 2, and~5, respectively) and progressively reduced for the denser spectral region above 40~Hz (\cref{tab:trunc_freq_damp,tab:trunc_mac}). This behaviour is physically interpretable: closely spaced or weakly excited modes impose greater demands, and the equidistant initial pole placement adopted here may not provide sufficient resolution in congested frequency bands. Adaptive or frequency-weighted pole initialisation strategies represent a natural and tractable extension. Here, LSCE fails to recover a majority of the modes, with MAC values below 0.04 for the four modes it does identify, highlighting its unsuitability for complex MIMO systems of this scale.

All things considered, these results establish FRVF as an efficient and noise-tolerant alternative for GVT-based modal analysis, competitive with iLF in large datasets.

\section{CONCLUSIONS}

FRVF is successfully adapted to accommodate non-square MIMO structural vibration datasets through the use of enhanced input stacking. Numerical validation demonstrates its high accuracy and strong resilience to measurement noise. The experimental application on the Hawk T1A aircraft shows that the method can be efficiently applied in realistic GVT scenarios, with performance comparable to that of other techniques, reported in previous studies. 
These results indicate that the enhanced FRVF provides a reliable and efficient approach to modal analysis under experimental conditions. Future work could focus on the development of automated mode selection criteria and the implementation of dynamic initial pole placement to further improve convergence.

\section*{ACKNOWLEDGMENT}
 The dataset \href{https://orda.shef.ac.uk/articles/dataset/BAE_T1A_Hawk_Full_Structure_Modal_Test/24948549}{BAE T1A Hawk Full Structure Modal Test} has been retrieved from the University of Sheffield data repository ORDA. The authors thank the LVV at the University of Sheffield for making the data openly available.



\end{document}